\documentclass[preprint,superscriptaddress,nofootinbib]{revtex4-1}
\usepackage{graphicx}
\usepackage{float} 
\usepackage{subfigure}
\usepackage{dcolumn}
\usepackage{bm}
\usepackage{amsmath}
\usepackage{amsfonts} 
\usepackage{latexsym}
\usepackage{bbm}
\usepackage{color}
\usepackage{amssymb}
\usepackage{amsthm}
\usepackage{epsf}
\usepackage{epsfig}
\usepackage{caption3}
\usepackage{appendix}
\usepackage{cases}
\usepackage{subfigure}
\usepackage{multirow}
\usepackage{hyperref}
\makeatletter

\newcommand{\Rmnum}[1]{\expandafter\@slowromancap\romannumeral #1@}
\makeatother

\newcommand{\be}{\begin{equation}}
	\newcommand{\ee}{\end{equation}}
\newcommand{\ba}{\begin{array}}
	\newcommand{\ea}{\end{array}}
\newcommand{\bea}{\begin{eqnarray}}
	\newcommand{\eea}{\end{eqnarray}}
\newcommand{\nn}{\nonumber}
\usepackage{theorem}
{
	\theoremheaderfont{\bfseries}
	\theorembodyfont{\normalfont}
}

\begin{document}

\title{Gravitational waves  from axions annihilation through quantum field theory}
\author{Jing Yang}

\author{Fa Peng Huang}%
\email{ Corresponding Author. huangfp8@sysu.edu.cn}

\affiliation{MOE Key Laboratory of TianQin Mission, TianQin Research Center for
	Gravitational Physics \& School of Physics and Astronomy, Frontiers
	Science Center for TianQin, Gravitational Wave Research Center of
	CNSA, Sun Yat-sen University (Zhuhai Campus), Zhuhai 519082, China}

\bigskip
	
	
\begin{abstract}
We use the scattering method to calculate the gravitational wave from axions annihilation in the axion cloud formed by the superradiance process around the Kerr black hole. We consider axions annihilating to gravitons as a three-body decay process and then calculate the corresponding decay width. In this approach, we can simply obtain the radiation power of gravitational wave and give the analytical approximate result with the spin effects of the Kerr black hole. Our study can also provide a cross-check to the numerical results in the traditional method.
\end{abstract}

\maketitle
	
\section{introduction}	
Motivated by the current situation of dark matter search, recently more and more studies are focused on the ultralight dark matter, such as ultralight axion or dark photon.  Axion~\cite{Peccei:1977hh} or axionlike particle~\cite{Svrcek:2006yi} is one of the  most promising dark matter candidates~\cite{Preskill:1982cy,Sikivie:2006ni}\footnote{In the following discussions, for simplicity, we use axion to represent axion and axionlike particle.}. 
As  pseudoscalar particles, ultralight axion particles can form Bose-Einstein condensation and further form dense axion object, like 
axion star, axion minicluster or axion cluster.  Especially, the so-called axion cloud can be formed from the superradiance process around Kerr black hole (BH)~\cite{Penrose:1969pc, zeld, Starobinsky:1973aij, Detweiler:1980uk}. 
Once the axion cloud is formed, abundant signals could be produced.
Especially, the axion cloud can be a promising gravitational wave (GW) source~\cite{Brito:2015oca,Arvanitaki:2014wva,Arvanitaki:2016qwi,Baryakhtar:2017ngi,Dev:2016hxv,Xie:2022uvp}, which may be detected by various GW experiments, such as LIGO/Virgo~\cite{LIGOScientific:2014pky,VIRGO:2014yos}, LISA~\cite{LISA:2017pwj}, TianQin~\cite{TianQin:2015yph}, and Taiji~\cite{Hu:2017mde}. These GW signals can be used to explore the properties of axion particles~\cite{Brito:2017zvb}. To unravel the axion properties from GW signals, it is crucial to obtain a reliable prediction of the GW radiation power from the axion cloud. 
Arvanitaki and Dubovsky~\cite{Arvanitaki:2010sy} studied the GW radiation power in the flat space approximation.
Brito et al.~\cite{Brito:2014wla} calculated the GW radiation power of axions annihilating to gravitons in the Schwarzschild metric.
Yoshino et al.~\cite{Yoshino:2013ofa} and Brito et al.~\cite{Brito:2017zvb} calculated  the GW radiation power in the Kerr metric.
However, the calculations of the GW radiation power through the traditional method are complicated in the Kerr metric. Due to the complicated numerical calculations in the Kerr metric,  the analytical approximation result has not been obtained.
To simplify the calculations and cross-check these important results, we use the method of scattering amplitude to calculate GW radiation power  generated from the axions annihilation process. As a by-product, we naturally obtain the analytical approximated result, which is convenient for the phenomenological study of GW search of ultralight axion particles at LIGO/Virgo, LISA, TianQin, and Taiji.

In Sec.~\ref{sec2}, we simply review the basic concepts about axion cloud from superradiance process, then we demonstrate the relevant Feynman rules for axions annihilating to gravitons and derive the formulations of the decay width. The radiation power of GWs from axions annihilating to gravitons is calculated in Sec.~\ref{sec3} where we give both analytical and numerical results. In Sec.~\ref{dis}, we derive the formula used in the flat space approximation through the scattering amplitude method in the classical source limit. Concise conclusions are given in Sec.~\ref{conc}.

\section{Scattering amplitude of axions annihilating to gravitons around Kerr black hole} \label{sec2}
\subsection{Axion cloud from superradiance around Kerr black hole}
When the Compton wavelength of the axion boson becomes comparable with the Kerr BH size,
axion cloud could form by extracting the energy and angular momentum of the Kerr BH through the so-called superradiance process.
In analogy with the atom, the axion cloud and the Kerr BH can form the ``gravitational atom".
This formation of axion cloud can be understood from the particle or wave aspect.
From the perspective of particle, the axion cloud can be formed due to the Penrose process~\cite{Penrose:1969pc} of Kerr BH since
there exists an ergosphere between the outer horizon and outer static limit surface.
From the perspective of wave, we could directly solve the Klein-Gordon equation in the Kerr metric.
At large distance, the radial solution is just a confluent hypergeometric function.
While, at small distance (close to the outer event horizon), the radial solution can be expressed as
the hypergeometric function. Matching them together, one can find the exponential growth solution when 
the Compton length of the axion is larger than the BH size~\cite{Detweiler:1980uk}. Besides the matched
	asymptotic method, the Klein-Gordon equation can also be exactly solved through the numerical methods~\cite{Dolan:2007mj}.
Once the axion cloud or gravitational atom is formed, it can emit GW signals from axions annihilating to gravitons or energy-level transitions
of the gravitational atom.
In this work, we focus on the GW signals from axions annihilating to gravitons.

\subsection{Relevant Feynman rules for axions annihilating to  gravitons}
The Einstein-Hilbert action in the general relativity is
\be
{\cal{S}}=\int d^4x {\cal{L}}=\int d^4 x \sqrt{-g}(\frac{2R}{\kappa^2}+{\cal L}_{\text{matter}}),
\ee
where the gravitational coupling $\kappa^2\equiv32\pi G=32\pi /M_{\text{pl}}^2$ with Planck mass $M_{\text{pl}}=1.22\times 10^{19}~\text{GeV}$, $g$ is the determinant of the metric field $g_{\mu\nu}$ and Riemann tensor  $R^{\mu}_{\nu\alpha\beta}\equiv\partial_{\alpha}\Gamma^{\mu}_{\nu\beta}-\partial_{\beta}\Gamma^{\mu}_{\nu\alpha}+\Gamma^{\mu}_{\sigma\alpha}\Gamma^{\sigma}_{\nu\beta}-\Gamma^{\mu}_{\sigma\beta}\Gamma^{\sigma}_{\nu\alpha}$.
We expand the metric as 
\be
g_{\mu \nu}=\eta_{\mu \nu}+\kappa h_{\mu \nu},
\ee
where $\eta_{\mu\nu}=\text{diag}\{1,-1,-1,-1\}$ is the flat space-time metric and $h_{\mu\nu}$ is the graviton field. Then we can get the expansion for the upper metric field $g_{\mu\nu}$ and $\sqrt{-g}$,
\bea
g^{\mu\nu}&=&\eta^{\mu\nu}-\kappa h^{\mu\nu}+\kappa^2 h^{\mu}_{\alpha}h^{\alpha\nu}+..., \nn \\
\sqrt{-g}&=&1+\frac{1}{2}\kappa h-\frac{1}{4}\kappa^2 h^{\alpha}_{\beta}h^{\beta}_{\alpha}+\frac{1}{8}\kappa^2h^2+...,
\eea
where $h=\eta^{\mu\nu}h_{\mu\nu} $ and we use $\eta^{\mu\nu}$ to raise and lower indices. The relevant Lagrangian of the axion field is ${\cal L}=\frac12 g_{\mu\nu}\partial^{\mu}\Phi\partial^{\nu}\Phi-\frac12 m_a^2 \Phi^2$, where $m_a$ is the axion mass and $\Phi$ represents the axion field.
After expanding the Einstein-Hilbert action with $h_{\mu\nu}$, one can get the following Feynman rules for axions annihilating to gravitons in the harmonic gauge \cite{DeWitt:1967uc, Bjerrum-Bohr:2004qcf}.

\begin{minipage}[h]{0.4\linewidth}\vspace{0.4cm}
	\centering\includegraphics[scale=0.27]{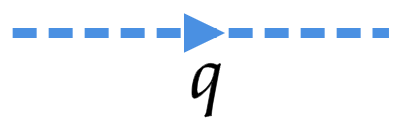}
\end{minipage}
\begin{minipage}[h]{0.65\linewidth}
	\centering$\displaystyle = \frac i{q^2-m_a^2+i\epsilon}$
\end{minipage}

\begin{minipage}[h]{0.4\linewidth}\vspace{0.4cm}
	\centering\includegraphics[scale=0.24]{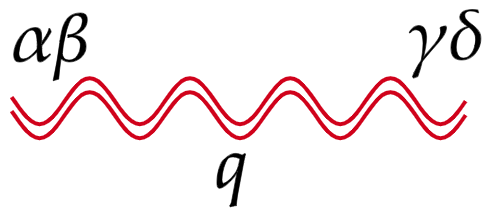}
\end{minipage}
\begin{minipage}[h]{0.59\linewidth}
	\centering $\displaystyle = \frac {i{\cal P}^{\alpha\beta\gamma\delta}}{q^2+i\epsilon}$
\end{minipage}\\
where
\begin{equation}
	{\cal P}^{\alpha\beta\gamma\delta} = \frac12\left(\eta^{\alpha\gamma}\eta^{\beta\delta} +
	\eta^{\beta\gamma}\eta^{\alpha\delta}
	-\eta^{\alpha\beta}\eta^{\gamma\delta}\right).
\end{equation}

\begin{minipage}[h]{0.4\linewidth}\vspace{0.15cm}
	\centering\includegraphics[scale=0.16]{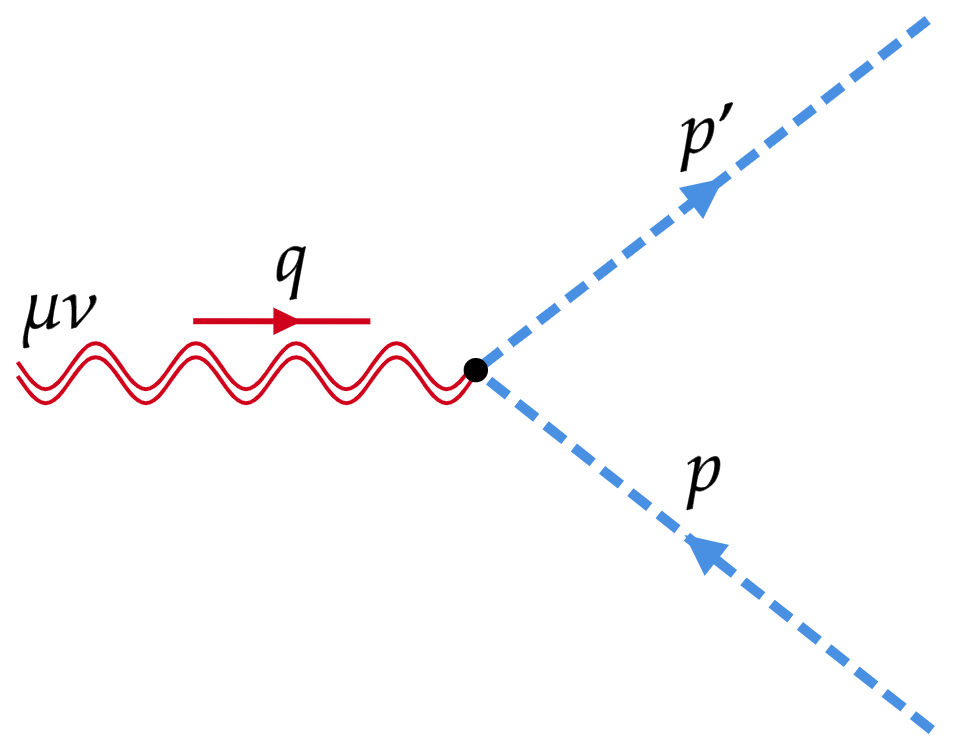}
\end{minipage}
\begin{minipage}[h]{0.65\linewidth}
	\centering$\displaystyle = \tau_1^{\mu \nu}(p,p',m_a)$
\end{minipage}\vspace{0.3cm}
where
\begin{equation}
	\tau_1^{\mu\nu}(p,p',m_a) = -\frac{i\kappa}2\Big (p^\mu p^{\prime \nu}
	+p^\nu p^{\prime \mu} - \eta^{\mu\nu}\left(p\cdot
	p^\prime-m_a^2\right)\Big).
\end{equation}

\begin{minipage}[h]{0.4\linewidth}\vspace{0.15cm}
	\centering\includegraphics[scale=0.15]{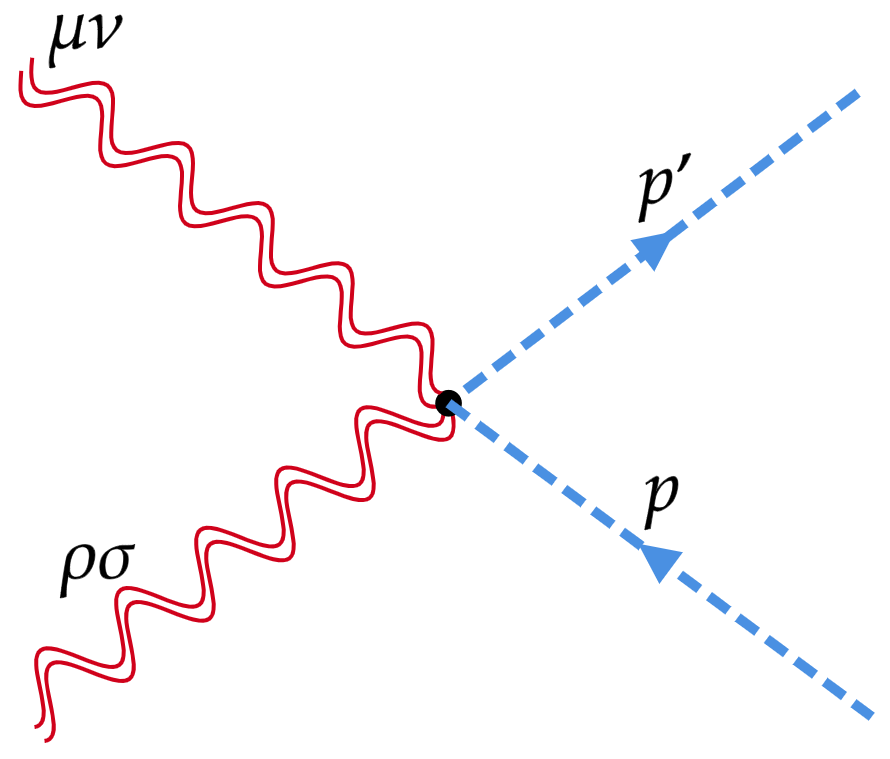}
\end{minipage}
\begin{minipage}[h]{0.65\linewidth}
	\centering$\displaystyle = \tau_2^{\mu \nu\rho\sigma}(p,p',m_a)$
\end{minipage}\vspace{0.3cm}
where
\begin{equation}{\begin{aligned}
			\tau_2^{\mu \nu \rho \sigma}(p,p')&= {i\kappa^2} \bigg (
			\Big ( I^{\mu \nu\alpha \delta} {I}^{\rho \sigma\beta}_{\ \
				\ \ \delta} - \frac14\left(\eta^{\mu \nu} I^{\rho
				\sigma\alpha \beta} +
			\eta^{\rho \sigma} I^{\mu \nu\alpha \beta} \right )
			\Big ) (p_\alpha p^\prime_{\beta} + p^\prime_{\alpha} p_\beta )  \\
			&-\frac12 \Big ( I^{\mu \nu\rho \sigma} - \frac12\eta^{\mu \nu}\eta^{\rho \sigma} \Big ) \Big ( p\cdot p' - m_a^2
			\Big)\bigg),\end{aligned}}
\end{equation}
with \bea
I_{\mu\nu\alpha\beta} &=& {1\over 2}
(\eta_{\mu\alpha}\eta_{\nu\beta}+\eta_{\nu\alpha}\eta_{\mu\beta}).
\eea

\begin{minipage}[h]{0.4\linewidth}\vspace{0.15cm}
	\centering\includegraphics[scale=0.14]{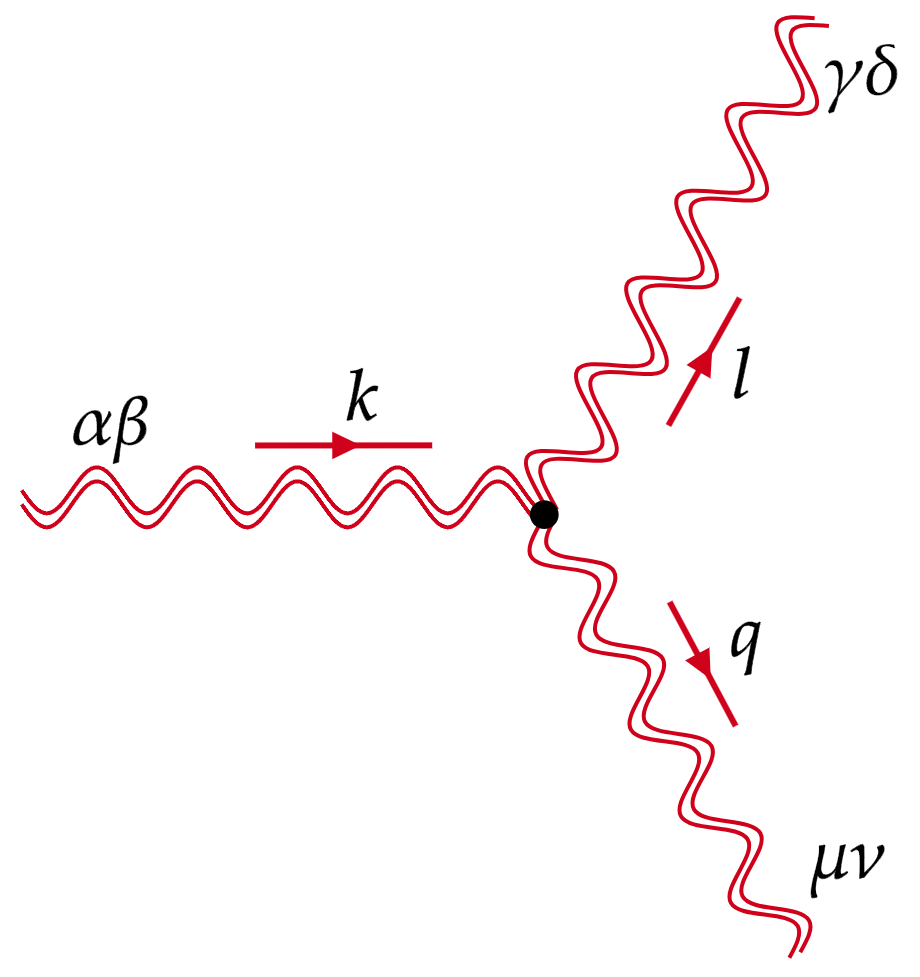}
\end{minipage}
\begin{minipage}[h]{0.65\linewidth}
	\centering$\displaystyle =
	{\tau_{3}}_{\alpha\beta\gamma\delta}^{\mu\nu}(k,q)$
\end{minipage}\vspace{0.3cm}
where
\begin{equation}{
		\begin{aligned}
			{\tau_{3}}_{\alpha  \beta \gamma \delta }^{\mu
				\nu}(k,q)&=-\frac{i\kappa}2\times \Bigg({\cal P}_{\alpha \beta
				\gamma \delta }\bigg(k^\mu k^\nu+ (k-q)^\mu (k-q)^\nu +q^\mu
			q^\nu-
			\frac32\eta^{\mu \nu}q^2\bigg)\\[0.00cm]&
			+2q_\lambda q_\sigma\Big( I_{\alpha \beta }^{\ \ \
				\sigma\lambda}I_{\gamma \delta }^{\ \ \ \mu \nu} + I_{\gamma
				\delta }^{\ \ \ \sigma\lambda}I_{\alpha \beta }^{\ \ \ \mu \nu}
			-I_{\alpha \beta }^{\ \ \ \mu  \sigma} I_{\gamma \delta }^{\ \ \
				\nu \lambda} - I_{\gamma \delta }^{\ \ \ \mu \sigma} I_{\alpha
				\beta }^{\ \ \ \nu \lambda}
			\Big)\\[0cm]&
			+\Big(q_\lambda q^\mu \big(\eta_{\alpha \beta }I_{\gamma \delta
			}^{\ \ \ \nu \lambda}+\eta_{\gamma \delta }I_{\alpha \beta }^{\ \
				\ \nu \lambda}\big) +q_\lambda q^\nu \big(\eta_{\alpha \beta
			}I_{\gamma \delta }^{\ \ \ \mu \lambda}+\eta_{\gamma \delta
			}I_{\alpha \beta }^{\ \ \ \mu  \lambda}\big)\\&
			-q^2\left(\eta_{\alpha \beta }I_{\gamma \delta }^{\ \ \ \mu
				\nu}-\eta_{\gamma \delta }I_{\alpha \beta }^{\ \ \ \mu \nu}\right)
			-\eta^{\mu \nu}q_\sigma q_\lambda\left(\eta_{\alpha \beta
			}I_{\gamma \delta }^{\ \ \ \sigma\lambda} +\eta_{\gamma \delta
			}I_{\alpha \beta }^{\ \ \
				\sigma\lambda}\right)\Big)\\[0cm]&
			+\Big(2q_\lambda\big(I_{\alpha \beta }^{\ \ \
				\lambda\sigma}I_{\gamma \delta \sigma}^{\ \ \ \ \nu}(k-q)^\mu
			+I_{\alpha \beta }^{\ \ \ \lambda\sigma}I_{\gamma \delta
				\sigma}^{\ \ \ \ \mu }(k-q)^\nu -I_{\gamma \delta }^{\ \ \
				\lambda\sigma}I_{\alpha \beta \sigma}^{\ \ \ \ \nu}k^\mu
			-I_{\gamma \delta }^{\ \ \ \lambda\sigma}I_{\alpha \beta
				\sigma}^{\ \ \ \ \mu }k^\nu \big)\\& +q^2\left(I_{\alpha \beta
				\sigma}^{\ \ \ \ \mu }I_{\gamma \delta }^{\ \ \ \nu \sigma} +
			I_{\alpha \beta }^{\ \ \ \nu \sigma}I_{\gamma \delta \sigma}^{\ \
				\ \ \mu }\right) +\eta^{\mu \nu}q_\sigma q_\lambda\big(I_{\alpha
				\beta }^{\ \ \ \lambda\rho}I_{\gamma \delta  \rho}^{\ \ \ \
				\sigma} +I_{\gamma \delta }^{\ \ \ \lambda\rho}I_{\alpha \beta
				\rho}^{\ \ \ \
				\sigma}\big)\Big)\\[0cm]&
			+\big(k^2+(k-q)^2\big)\bigg(I_{\alpha \beta }^{\ \ \ \mu
				\sigma}I_{\gamma \delta \sigma}^{\ \ \ \ \nu} +I_{\gamma \delta
			}^{\ \ \ \mu  \sigma}I_{\alpha \beta \sigma}^{\ \ \ \ \nu}
			-\frac12\eta^{\mu \nu}{\cal P}_{\alpha \beta \gamma \delta
			}\bigg)-I_{\gamma \delta }^{\ \ \ \mu \nu}\eta_{\alpha
				\beta }k^2-I_{\alpha \beta }^{\ \ \ \mu \nu}\eta_{\gamma \delta
			}(k-q)^2\Bigg).
	\end{aligned}}
\end{equation}

\subsection{Formulations of decay width }
The annihilation process of two axions near the BH can be understood as a decay process of a three-body bound state which is composed of two axions and a BH.  
 If the “fine structure constant” $\alpha=GM_bm_a$ of the gravitational system is small where $M_b$ is the BH mass, then the internal motions are slow and we can treat the bound state as a nonrelativistic system. Thus we can approximately expand the bound state which is denoted as $|B\rangle$ in states of the free field theory in the center-of-mass (COM) frame as
\be
|B\rangle=\sqrt{2M}\iint \frac{d^3\boldsymbol{p_1}}{(2\pi)^3}\frac{d^3\boldsymbol{p_2}}{(2\pi)^3} \Psi_1(\boldsymbol{p_1})\Psi_2(\boldsymbol{p_2})\frac{1}{\sqrt{2E_1}\sqrt{2E_2}\sqrt{2E_b}}|\boldsymbol{p_b},\boldsymbol{p_1},\boldsymbol{p_2}\rangle ,
\ee
where $\boldsymbol{p_b}$ is the momentum of BH, $\boldsymbol{p_1}$, $\boldsymbol{p_2}$ are the momenta of two axions satisfying $\boldsymbol{p_b}=-\boldsymbol{p_1}-\boldsymbol{p_2}$ in the COM frame, $\Psi_1$($\Psi_2$) is the wave function of bounded axion in momentum space, and $M$ is the mass of the bound state. $E_1$,~$E_2$,~$E_b$ are the energy of axions and BH respectively. 

\begin{figure}[!htp]
	\centering  
	\subfigure[]{
		\label{1a}
		\includegraphics[width=0.4\textwidth]{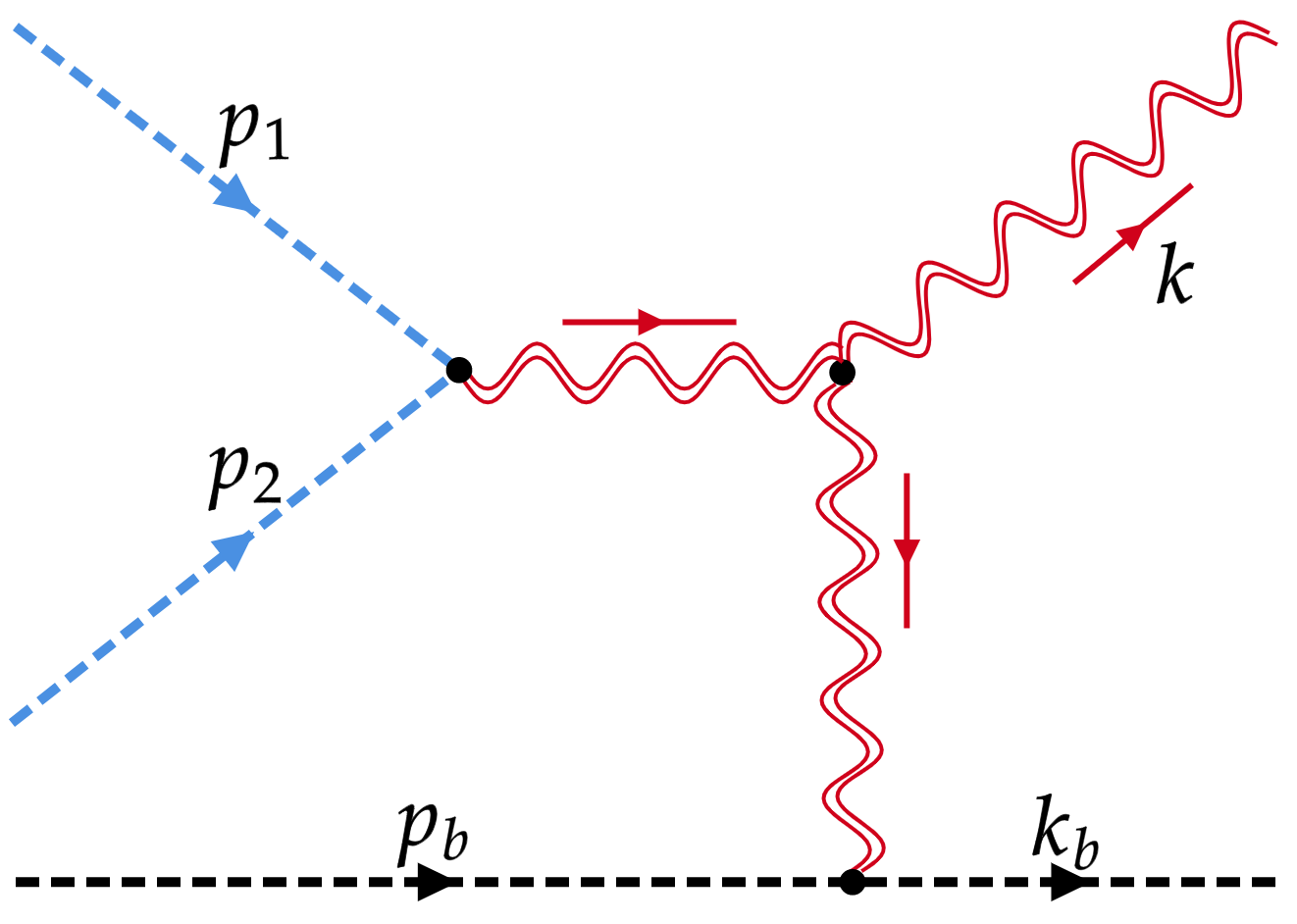}}
	\subfigure[]{
		\label{1b}
		\includegraphics[width=0.39\textwidth]{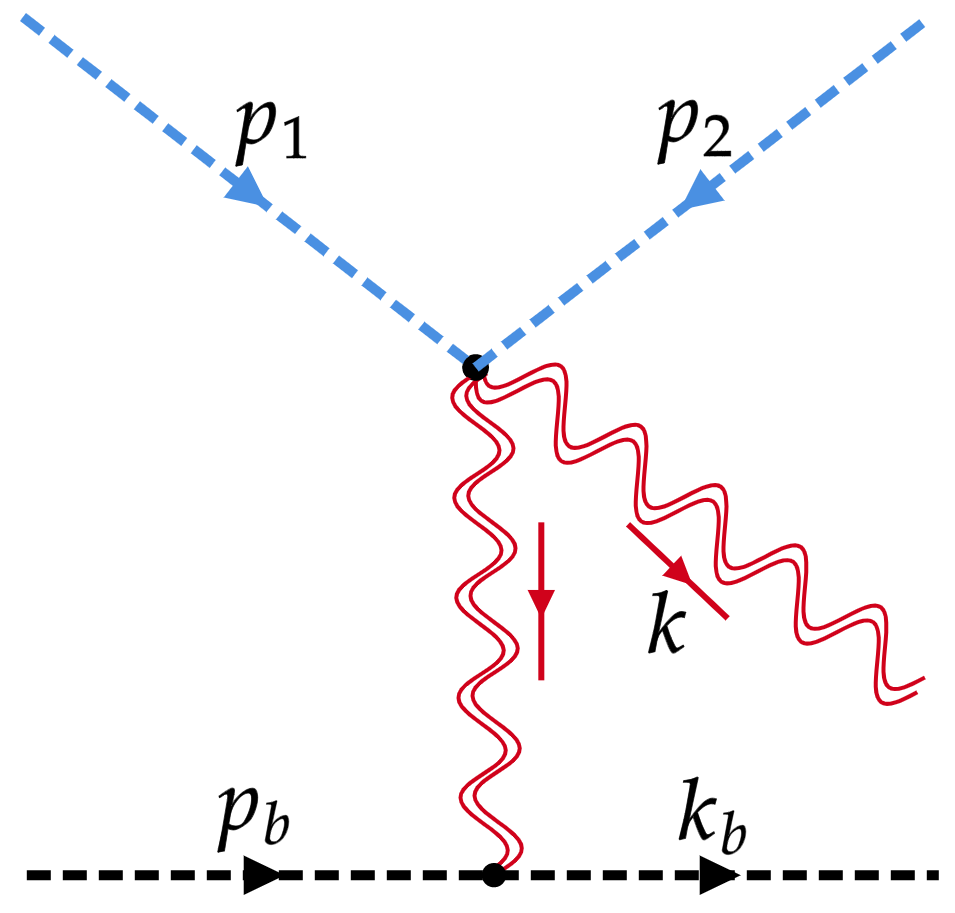}}
	\centering  
	\ \ \subfigure[]{
		\label{1c}
		\includegraphics[width=0.4\textwidth]{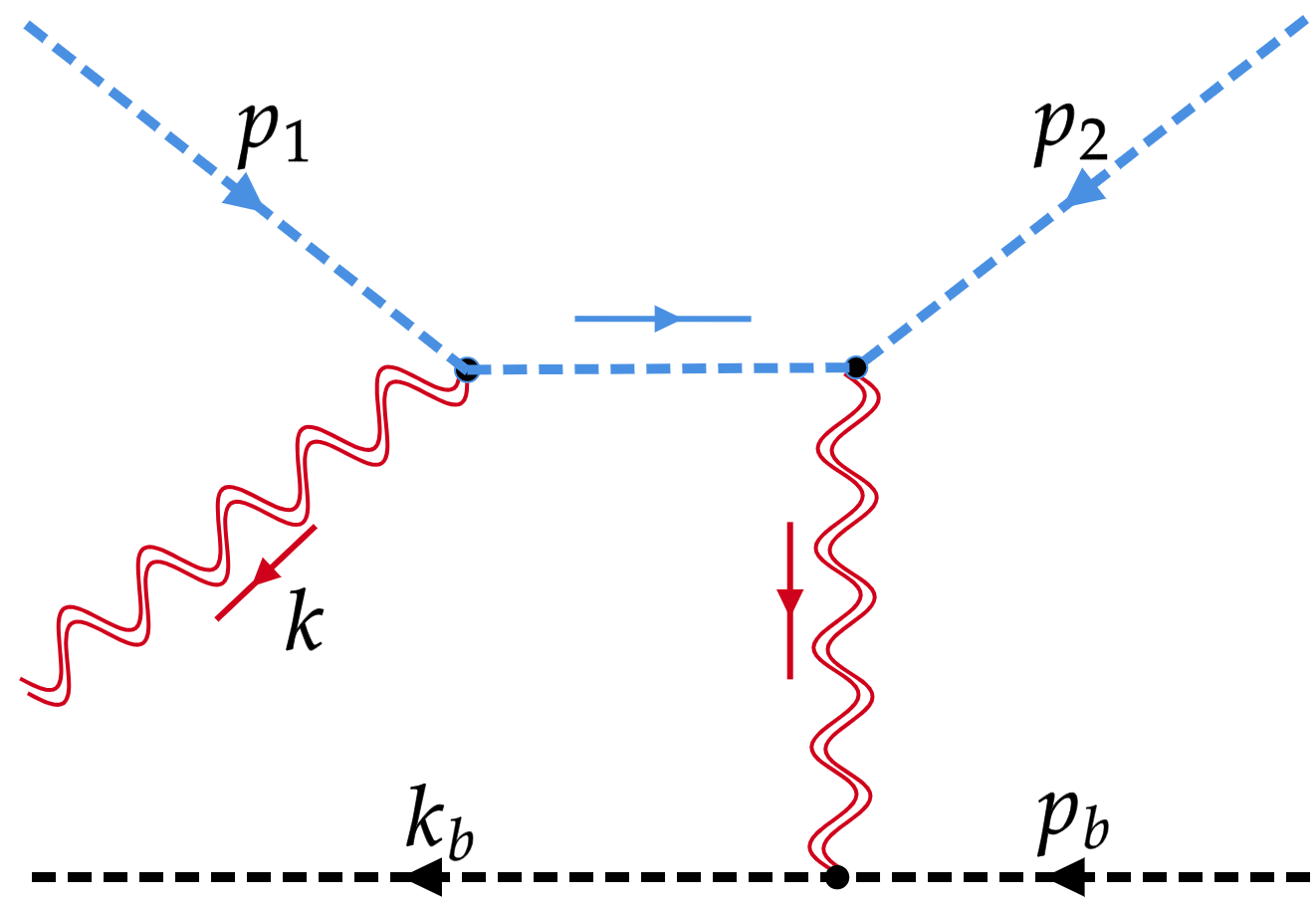}}
	\subfigure[]{
		\label{1d}
		\includegraphics[width=0.4\textwidth]{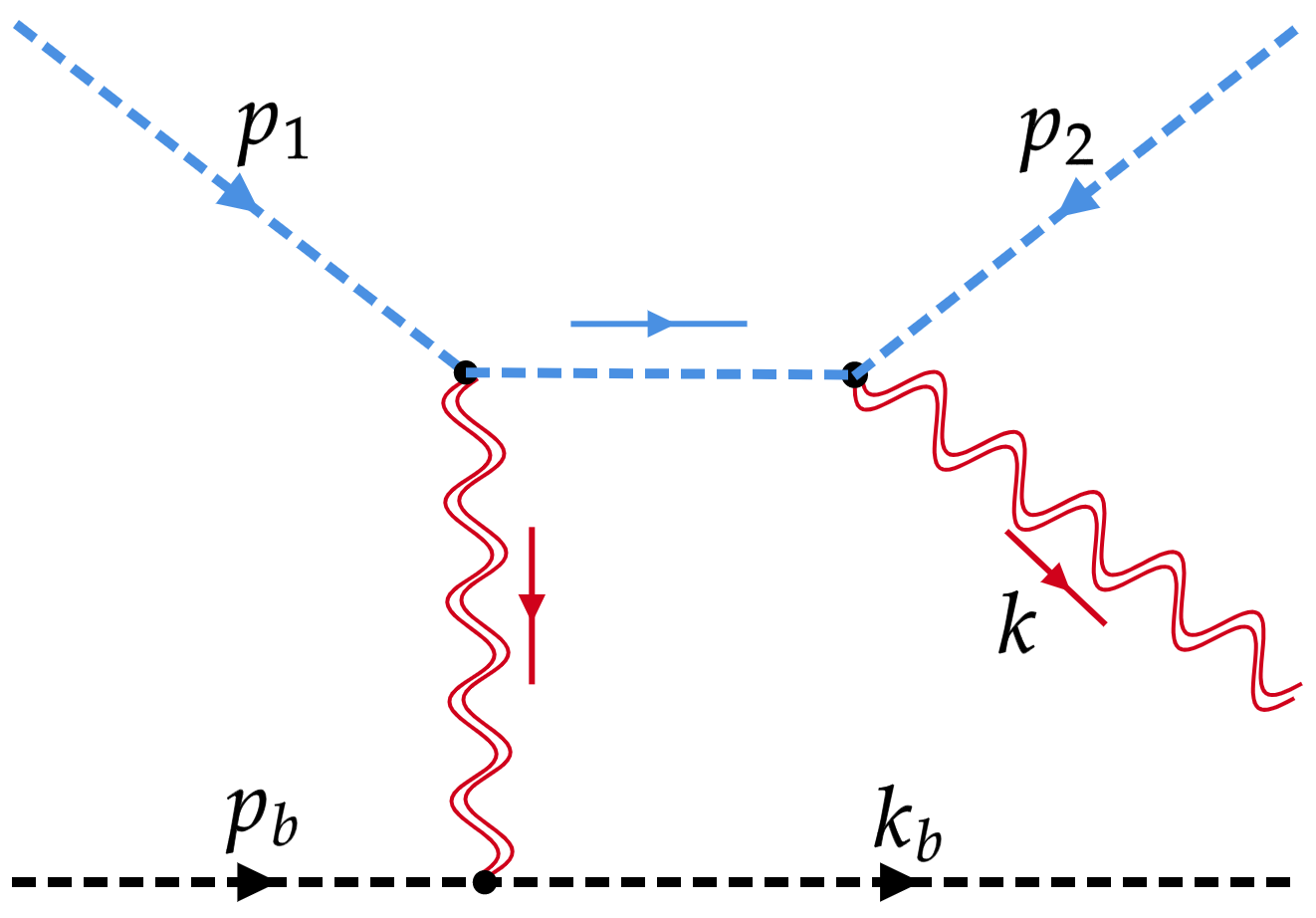}}
	\caption{The Feynman diagrams for the scattering amplitude of $|\boldsymbol{p_b},\boldsymbol{p_1},\boldsymbol{p_2}\rangle \rightarrow |\boldsymbol{k},\boldsymbol{k_b}\rangle $ process. The blue and black dashed lines depict the axion and BH respectively while the red double-wavy line represents the graviton.    } \label{fd}
\end{figure}

The decay with of the bound state in the COM frame is
\bea\label{gamma}
\Gamma&\simeq&\iint \frac{d^3\boldsymbol{k}}{(2\pi)^3}\frac{d^3\boldsymbol{k_b}}{(2\pi)^3}\frac{1}{2|\boldsymbol{k}|}\frac{1}{2E_{b}'}\big{|}\iint \frac{d^3\boldsymbol{p_1}}{(2\pi)^3}\frac{d^3\boldsymbol{p_2}}{(2\pi)^3} \Psi_1(\boldsymbol{p_1})\Psi_2(\boldsymbol{p_2})\frac{1}{\sqrt{2E_1}\sqrt{2E_2}\sqrt{2E_b}} \nn \\
&&{\cal M}(\boldsymbol{p_b},\boldsymbol{p_1},\boldsymbol{p_2}\rightarrow \boldsymbol{k},\boldsymbol{k_b})\big{|}^2  (2\pi)^4\delta^4(p_1+p_2+p_b -k-k_b),
\eea
where  ${\cal M}(\boldsymbol{p_b},\boldsymbol{p_1},\boldsymbol{p_2}\rightarrow \boldsymbol{k},\boldsymbol{k_b})$ is the scattering amplitude of $|\boldsymbol{p_b},\boldsymbol{p_1},\boldsymbol{p_2}\rangle \rightarrow |\boldsymbol{k},\boldsymbol{k_b}\rangle $ process, $\boldsymbol{k}$ and $\boldsymbol{k_b}$ are the momenta of graviton and BH respectively in the final state, $E_b'$ is the energy of BH in the final state. There are four Feynman diagrams relevant to the annihilation process $a+a+b\rightarrow g+b$ at tree level as shown in Fig.~\ref{fd} where $a$, $b$ and $g$ represent the axion, BH and graviton respectively.

We can approximately take the COM frame as the BH frame due to $M_b \gg m_a$, which means we can take the static source limit
\be
E_b\simeq E_b'\simeq M_b, \ \ \ \boldsymbol{p_b}=0 .
\ee
Then we have
\bea
\Gamma&\simeq&\int\frac{d^3\boldsymbol{k}}{(2\pi)^3}\frac{1}{2|\boldsymbol{k}|}\frac{1}{2M_b}\big{|}\iint \frac{d^3\boldsymbol{p_1}}{(2\pi)^3}\frac{d^3\boldsymbol{p_2}}{(2\pi)^3} \Psi_1(\boldsymbol{p_1})\Psi_2(\boldsymbol{p_2})\frac{1}{\sqrt{2E_1}\sqrt{2E_2}\sqrt{2M_b}} \nn \\
&&{\cal M}(\boldsymbol{p_b}=0,\boldsymbol{p_1},\boldsymbol{p_2}\rightarrow \boldsymbol{k},\boldsymbol{k_b}=-\boldsymbol{k})\big{|}^2 2\pi \delta(E_1+E_2-|\boldsymbol{k}|),
\eea 
where  ${\cal M}(\boldsymbol{p_b}=0,~\boldsymbol{p_1},\boldsymbol{p_2}\rightarrow \boldsymbol{k},\boldsymbol{k_b}=-\boldsymbol{k})$ is just the scattering amplitude of two axions annihilating to a graviton in the static source background.

\begin{figure}[!htp]
	\begin{center}
		\includegraphics[width=0.7\textwidth]{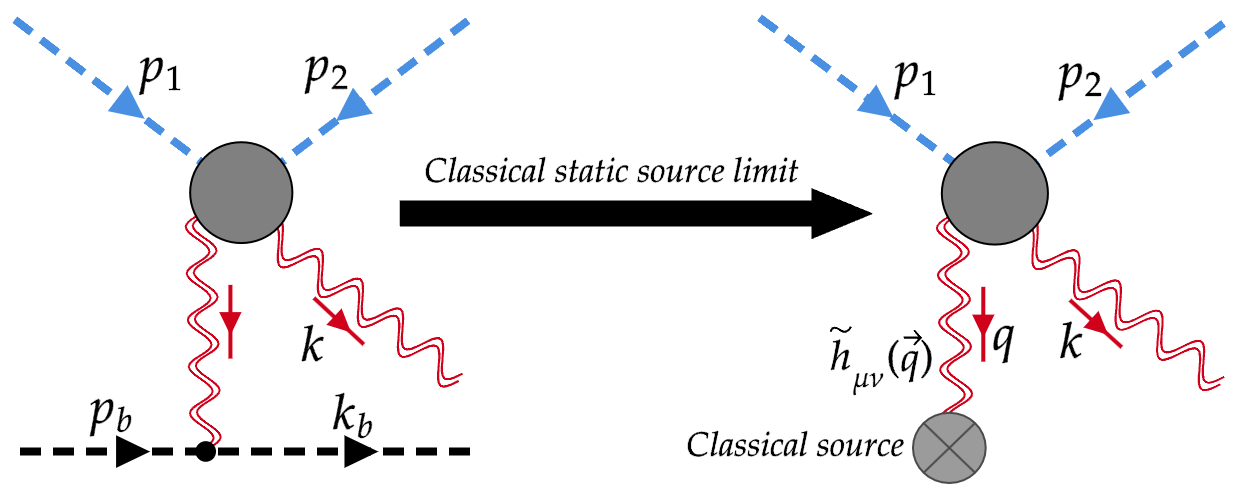}
		\caption{The Feynman diagram for the scattering amplitude of $|\boldsymbol{p_b},\boldsymbol{p_1},\boldsymbol{p_2}\rangle \rightarrow |\boldsymbol{k},\boldsymbol{k_b}\rangle $ process (left panel). We can take the classical static source limit due to $M_b\gg m_a$ which means the gravitational field generated by a BH is considered as a background field (right panel).   } \label{fig2}
	\end{center} 
\end{figure}

We can decompose the scattering amplitude of
$\langle  \boldsymbol{k},\boldsymbol{k_b}|\boldsymbol{p_b},\boldsymbol{p_1},\boldsymbol{p_2}\rangle $ into 

\begin{minipage}[h]{0.29\linewidth}
	\centering$\displaystyle i{\cal M}(\boldsymbol{p_b},\boldsymbol{p_1},\boldsymbol{p_2}\rightarrow \boldsymbol{k},\boldsymbol{k_b}) =$
\end{minipage}\vspace{0.1cm}
\begin{minipage}[h]{0.3\linewidth}\vspace{0.1cm}
	\centering\includegraphics[scale=0.3]{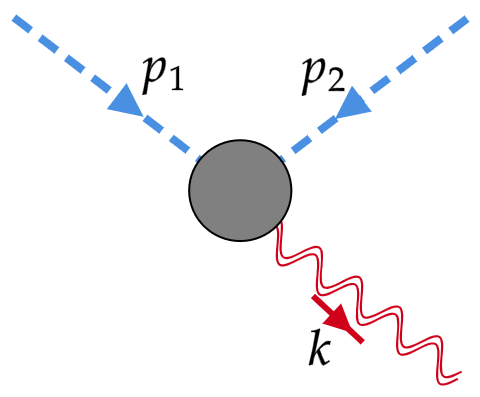}
\end{minipage}
\begin{minipage}[h]{0.05\linewidth}
	\centering$\displaystyle \times$
\end{minipage}\vspace{0.1cm}
\begin{minipage}[h]{0.3\linewidth}\vspace{0.1cm}
	\centering\includegraphics[scale=0.3]{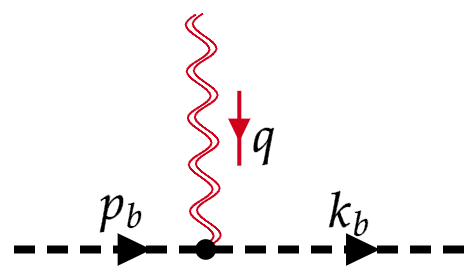}
\end{minipage}
\be \label{sc}
\ \ \ =i{\cal M}_1^{\mu \nu}(\boldsymbol{p_1},\boldsymbol{p_2}, \boldsymbol{k})\times {\cal M}_{2\mu \nu}(\boldsymbol{p_b},\boldsymbol{k_b}),\ \ \ \ \ \ 
\ee

where
\bea
(2\pi)^4 \delta^4(k_b-p_b-q){\cal M}_{2\mu\nu}&=&\frac{\kappa{\cal P}_{\mu\nu}^{\alpha\beta}\int d^4 xe^{-iq\cdotp x}\langle \boldsymbol{k_b}|{\cal T}_{b\alpha\beta}|\boldsymbol{p_b}\rangle }{2(k_b-p_b)^2} \nn \\
&=&\frac{\kappa\int d^4 xe^{-iq\cdotp x}\langle \boldsymbol{k_b}|{\cal T}_{b\mu\nu}-\eta_{\mu\nu}{\cal T}_b/2|\boldsymbol{p_b}\rangle }{2(k_b-p_b)^2}.
\eea
Here ${\cal T}_{b\mu\nu}$ is energy-momentum tensor of the BH's field and ${\cal T}_b={\cal T}^{\mu}_{b\mu}$. We define ${\cal T}_{b\mu\nu}={\cal T}'_{b\mu\nu}e^{i(k_b-p_b)\cdotp x}$ because $\langle \boldsymbol{k_b}|{\cal T}_{b\alpha\beta}|\boldsymbol{p_b}\rangle $ has a factor $e^{i(k_b-p_b)\cdotp x}$. Then we find
\be 
{\cal M}_{2\mu\nu}=\frac{\kappa\langle \boldsymbol{k_b}|{\cal T}'_{b\mu\nu}-\eta_{\mu\nu}{\cal T}'_b/2|\boldsymbol{p_b}\rangle }{2(k_b-p_b)^2}.
\ee
In the above discussion we only consider the local interaction between massive particles and gravitons in the quantum field theory. While for a BH, it has continuously distributed energy-momentum in a region, namely, the interactions take place in the corresponding region. So  we need to integrate over the whole interaction region by extending the definition of the energy-momentum tensor in the quantum field theory as
\be
{\cal T}_{b\mu\nu}=\int d^3 \boldsymbol{y} \tilde{{\cal T}}_{b\mu\nu}(\boldsymbol{y})e^{i(k_b-p_b)\cdotp x}e^{-i(\boldsymbol{k_b-p_b})\cdotp \boldsymbol{y}}.
\ee
$\tilde{{\cal T}}_{b\mu\nu}(\boldsymbol{y})$ is the energy-momentum density operator, $\boldsymbol{y}$ is the position relative to $\boldsymbol{x}$, and the factor $e^{-i(\boldsymbol{k_b-p_b})\cdotp \boldsymbol{y}}$ is understood as a translation operator relative to $\boldsymbol{x}$.
Thus the relation between energy-momentum tensor of BH $T_{b\mu \nu}$ and $\tilde{{\cal T}}_{b\mu\nu}$ is
\be
\langle \boldsymbol{p}_b|\tilde{{\cal T}}_{b\mu\nu}(\boldsymbol{y})|\boldsymbol{p}_b\rangle \xrightarrow{\text{classical static limit}}2M_b T_{b\mu \nu}(\boldsymbol{y}),
\ee
where the factor $2M_b$ comes from the normalization of the state $\langle \boldsymbol{k}|\boldsymbol{p} \rangle=2E_{\boldsymbol{p}}(2\pi)^3\delta^3(\boldsymbol{p}-\boldsymbol{k})$ with $E_{\boldsymbol{p}}=\sqrt{\boldsymbol{p}^2+M_b^2}$ in the quantum field theory.
In the static limit, $|\boldsymbol{k_b}-\boldsymbol{p_b}|\ll M_b$, we have  
\be\langle \boldsymbol{k}_b|\tilde{{\cal T}}_{b\mu\nu}(\boldsymbol{y})|\boldsymbol{p}_b\rangle \approx \langle \boldsymbol{p}_b|\tilde{{\cal T}}_{b\mu\nu}(\boldsymbol{y})|\boldsymbol{p}_b\rangle .
\ee 
Thus 
\bea \label{M2}
{\cal M}_{2\mu\nu}&=&\frac{-\kappa  M_b\int d^3\boldsymbol{y}e^{-i(\boldsymbol{k_b-p_b})\cdotp \boldsymbol{y}} (T_{b\mu \nu}(\boldsymbol{y})-\eta_{\mu \nu }T(\boldsymbol{y})/2)}{(\boldsymbol{k}_b-\boldsymbol{p}_b)^2} \nn \\
&=&\frac{-\kappa  M_b (T_{b\mu \nu}(\boldsymbol{q})-\eta_{\mu \nu }T(\boldsymbol{q})/2)}{\boldsymbol{q}^2}.
\eea
Note that the Einstein equation in the weak-field approximation is
\be
\Box h_{\mu \nu}=-\frac{\kappa}{2}(T_{\mu\nu}-\eta_{\mu\nu}T/2),
\ee
if the source is static, we have 
\be
\nabla^2 h_{\mu \nu}=\frac{\kappa}{2}(T_{\mu\nu}-\eta_{\mu\nu}T/2),
\ee
then the solution reads
\be
h_{\mu \nu}(\boldsymbol{x})=-\frac{\kappa}{8\pi}\int d^3 \boldsymbol{y}\frac{(T_{\mu\nu}(\boldsymbol{y})-\eta_{\mu\nu}T(\boldsymbol{y})/2)}{|\boldsymbol{x}-\boldsymbol{y}|}.
\ee
The Fourier transformation of $h_{\mu\nu}(\boldsymbol{x})$ in momentum space is
\be \label{hq}
\tilde{h}_{\mu \nu}(\boldsymbol{q})=-\frac{\kappa}{2}\frac{T_{b\mu \nu}(\boldsymbol{q})-\eta_{\mu \nu }T(\boldsymbol{q})/2}{\boldsymbol{q}^2}.
\ee
Substituting Eq.~\eqref{hq} into Eq.~\eqref{M2}, we get
\be \label{M21}
{\cal M}_{2\mu \nu}=2M_b\tilde{h}_{\mu \nu}(\boldsymbol{q}),
\ee
where $\tilde{h}_{\mu \nu}(\boldsymbol{q})$ is the field generated by the Kerr BH in momentum space. Thus the five-point amplitudes in Fig~\ref{fd} can be calculated as two axions annihilating in the background field $\tilde{h}_{\mu \nu}(\boldsymbol{q})$ as depicted in Fig~\ref{fig2}.
Then the decay width $\Gamma$ can be simplified to
\be
\Gamma=\int\frac{d^3\boldsymbol{k}}{(2\pi)^3}\frac{\pi \delta(E_1+E_2-|\boldsymbol{k}|)}{|\boldsymbol{k}|} 
\big{|}\iint \frac{d^3\boldsymbol{p_1}}{(2\pi)^3}\frac{d^3\boldsymbol{p_2}}{(2\pi)^3} \frac{\Psi_1(\boldsymbol{p_1})\Psi_2(\boldsymbol{p_2})}{\sqrt{2E_1}\sqrt{2E_2}}{\cal M}_1^{\mu\nu}(\boldsymbol{p_1},\boldsymbol{p_2},\boldsymbol{k})\tilde{h}_{\mu\nu}(\boldsymbol{q})\big{|}^2.
\ee
The gravitational background field of the Kerr BH in the weak-field approximation is
\be
h_{\mu \nu}(\boldsymbol{x})=h_{\mu \nu ,0}(\boldsymbol{x})+h_{\mu \nu ,J}(\boldsymbol{x}), 
\ee
\be\label{h0}
h_{\mu \nu ,0}(\boldsymbol{x})=N_{\mu \nu}\frac{\kappa M_b}{16\pi r},
\ee 
\be\label{hj}
 h_{\mu \nu ,J}(\boldsymbol{x})=-N^i_{\mu \nu}\frac{\kappa }{16\pi r^3}\epsilon^{ikm}x^kJ^m, 
\ee 
where
\be
N_{\mu \nu}=\eta_{\mu \nu}-2\eta_{\mu 0}\eta_{\nu 0},
\ee
\be
N^i_{\mu \nu}=\eta^{0}_{\mu}\eta^{i}_{\nu}+\eta^{0}_{\nu}\eta^{i}_{\mu}, \  i=1,2,3.
\ee

Then the gravitational background field of the Kerr BH in the momentum space $\tilde{h}_{\mu\nu}(\boldsymbol{q})$ can be  derived as (Note that the direction of $\boldsymbol{q}$ is incoming to the Kerr BH in the Feynman diagrams which is opposite to the usual Fourier transform convention where the direction of momentum is outgoing.)
\be \label{hqj}
\tilde{h}_{\mu\nu}(\boldsymbol{q})=\tilde{h}_{\mu\nu,0}(\boldsymbol{q})+\tilde{h}_{\mu\nu,J}(\boldsymbol{q}),
\ee
\be
\tilde{h}_{\mu\nu,0}(\boldsymbol{q})=N_{\mu\nu}\frac{\kappa M_b}{4\boldsymbol{q}^2},
\ee
\be
\tilde{h}_{\mu\nu,J}(\boldsymbol{q})=N^i_{\mu\nu}\frac{-i\kappa \epsilon^{ikm}q^kJ^m}{4\boldsymbol{q}^2}.
\ee

We have obtained the expression of ${\cal M}_2^{\mu\nu}$ in the static source limit above, and then we turn to the first piece ${\cal M}_1^{\mu\nu}$ in Eq.~\eqref{sc} which is the sum of the corresponding pieces of four Feynman diagrams in Fig.~\ref{fd},
\be
{\cal M}_1^{\mu\nu}=\sum_{s=\pm}({\cal M}_{1,a}^{s,\mu\nu}+{\cal M}_{1,b}^{s,\mu\nu}+{\cal M}_{1,c}^{s,\mu\nu}+{\cal M}_{1,d}^{s,\mu\nu}),
\ee
with
\be
{\cal M}_{1,a}^{s,\mu\nu}=\tau_1^{\rho\sigma}(p_1,-p_2)\frac{i{\cal P}^{\gamma \delta}_{\rho \sigma}}{(p_1+p_2)^2}\tau^{\mu\nu}_{3,\alpha\beta\gamma\delta}(-k,q)\epsilon^{s,\alpha \beta}\ \ \text{for Fig.~\ref{1a}},
\ee
\be
{\cal M}_{1,b}^{s,\mu\nu}=\tau_{2,\alpha \beta}^{\mu\nu}(p_1,-p_2)\epsilon^{s,\alpha \beta}\ \ \text{for Fig.~\ref{1b}},
\ee
\be
{\cal M}_{1,c}^{s,\mu\nu}=\tau_{1,\alpha\beta}(p_1,p_1-k)\frac{i}{(p_1-k)^2-m_a^2}\tau^{\mu\nu}_{1}(p_1-k,-p_2)\epsilon^{s,\alpha \beta}\ \ \text{for Fig.~\ref{1c}},
\ee
\be
{\cal M}_{1,d}^{s,\mu\nu}=\tau_{1}^{\mu\nu}(p_1,k-p_2)\frac{i}{(k-p_2)^2-m_a^2}\tau_{1,\alpha\beta}(k-p_2,-p_2)\epsilon^{s,\alpha \beta}\ \ \text{for Fig.~\ref{1d}},
\ee
where $\epsilon^{s,\alpha \beta}$ is the polarization tensor of the graviton.

The polarization tensor of the physical graviton can be described
in terms of a direct product of unit spin polarization vectors \cite{Bjerrum-Bohr:2014lea} 
\bea \label{epsilon}
\epsilon^{+}_{\mu \nu}&=&\epsilon^{+}_{\mu}\epsilon^{+}_{\nu},\ \ \ \text{for  helicity}=+2, \nn \\
\epsilon^{-}_{\mu \nu}&=&\epsilon^{-}_{\mu}\epsilon^{-}_{\nu},\ \ \ \text{for  helicity}=-2.
\eea
In the harmonic gauge, they satisfy
\be
\eta^{\mu \nu}\epsilon^{+}_{\mu}\epsilon^{+}_{\nu}=\eta^{\mu \nu}\epsilon^{-}_{\mu}\epsilon^{-}_{\nu}=0,\ \ k^{\mu}\epsilon^{\pm}_{\mu}=0.
\ee
We take the  spin direction of the BH as the $z$-axis, then the unit polarization vectors can be written as
\bea
\epsilon^{+}_{\mu}&=&\frac{1}{\sqrt{2}}(0,-\cos{\theta}\cos{\phi}+i \sin \phi ,-\cos \theta  \sin \phi -i \cos \phi ,\sin
\theta ), \nn \\
\epsilon^{-}_{\mu}&=&\frac{1}{\sqrt{2}}(0,\cos{\theta}\cos{\phi}+i \sin \phi ,\cos \theta  \sin \phi -i \cos \phi ,-\sin
\theta ),
\eea
where $\theta$ and $\phi$ are polar and azimuth angles of the momentum $\boldsymbol{k}$ in the spherical coordinate.

\section{radiation power of Gravitational wave } \label{sec3}

The superradiant rate of different energy level from the Kerr BH with spin angular momentum $J$ is estimated as \cite{Arvanitaki:2010sy}
\be
\Gamma_{n l m }=2 m_a \alpha^{4 l+4} (GM_b+\sqrt{G^2M_b^2-J^2/M_b^2})\left( \frac{m}{2G^2M_b^3}\frac{J}{1+
\sqrt{1-(J/GM_b^2)^2}}-m_{a}\right) C_{n l m },
\ee
where
\bea
C_{n l m }&=&\frac{2^{4 l+2}(l+n) !}{n^{2 l+4} (n-l-1) !}\left(\frac{l !}{(2 l) !(2 l+1) !}\right)^{2} \prod_{j=1}^{l}\Bigg(j^{2}\left(1-\frac{J^{2}}{G^{2}M_b^4}\right)\nn \\
&+& 4\left(GM_b+\sqrt{G^2M_b^2-J^2/M_b^2}\right)^{2}
\left( \frac{m}{2G^2M_b^3}\frac{J}{1+
\sqrt{1-(J/GM_b^2)^2}}-m_{a}\right)^2\Bigg).
\eea
 Thus the fastest superradiant level of axions in the cloud is $(n=2,l=1,m=1)$ where $n$,~$l$,~$m$ are the principal, orbital and magnetic quantum number. 

   In the following calculations, we can only consider the axions annihilating in the $(n=2,l=1,m=1)$ state of the cloud, and the wave function of the $(n=2,l=1,m=1)$ state can be obtained by solving the Schrödinger equation of the gravitational atom,
  \be
  \Psi (\boldsymbol{x})=\frac{\sqrt{6}r}{12 a_0^{5/2}}e^{-r/2a_0}\sqrt{\frac{3}{8\pi}}\sin \theta e^{i\phi},
  \ee
where $a_0=1/(\alpha m_a)$ is the Bohr radius of the gravitational atom and $r$ is the distance to BH. The wave function in momentum space can be derived as 
 \be
 \Psi(\boldsymbol{p})=-i 2\sqrt{\pi}a_0^{3/2}\frac{a_0(p_x+ip_y)}{(a_0^2\boldsymbol{p}^2+1/4)^3}.
 \ee
 
The axions annihilating to gravitons in the system of BH and axions cloud can be considered as the process of $N(N-1)/2$ three-body bound states decay where $N$ is the total number of axions in the state $(n=2,l=1,m=1)$. And the binding energy of the corresponding energy-level is estimated as $E_n=-\alpha^2m_a/(2n^2)$. The mass of the three-body bound state can be approximated as $M=M_b+2m_a+2E_n$. Thus the  radiation power can be calculated by
 \bea \label{P}
P&=&-\frac{dE}{dt}=\frac{N(N-1)}{2}(M-M_b)\Gamma
\simeq \frac{N^2}{2}(M-M_b)
\int\frac{d^3\boldsymbol{k}}{(2\pi)^3}\frac{\pi \delta(E_1+E_2-|\boldsymbol{k}|)}{|\boldsymbol{k}|} \nn \\
&&\big{|}\iint \frac{d^3\boldsymbol{p_1}}{(2\pi)^3}\frac{d^3\boldsymbol{p_2}}{(2\pi)^3} \frac{\Psi_1(\boldsymbol{p_1})\Psi_2(\boldsymbol{p_2})}{\sqrt{2E_1}\sqrt{2E_2}}{\cal M}_1^{\mu\nu}(\boldsymbol{p_1},\boldsymbol{p_2},\boldsymbol{k})\tilde{h}_{\mu\nu}(\boldsymbol{q})\big{|}^2.
\eea
The mass difference ($M-M_b$) of the three-body bound state ($M$) and BH ($M_b$) is  the energy of emitting graviton. For two axions at the energy-level $(n=2,l=1,m=1)$, $M-M_b=2m_a(1-\alpha^2/8)$.

 For the sake of $E_1\gg |\boldsymbol{p_1}|$ and $E_2\gg |\boldsymbol{p_2}|$ when $\alpha$ is small, we expand the propagators in the nonrelativistic limit to third order as
\bea
\frac{1}{(p_1+p_2)^2}&=&\frac{1}{2m_a^2+2E_1E_2-2\boldsymbol{p_1}\cdot\boldsymbol{p_2}}=\frac{1}{2}\Big(\frac{1}{m_a^2+E_1E_2}+\frac{\boldsymbol{p_1}\cdot\boldsymbol{p_2}}{(m_a^2+E_1E_2)^2} \nn \\
&+&\frac{(\boldsymbol{p_1}\cdot\boldsymbol{p_2})^2}{(m_a^2+E_1E_2)^3}+\cdots\Big),
\eea
\bea
\frac{1}{(p_i-k)^2-m_a^2}=\frac{1}{2}\frac{-1}{E_i|\boldsymbol{k}|-\boldsymbol{p_i}\cdot \boldsymbol{k}}=-\frac{1}{2}\Big (\frac{1}{E_i|\boldsymbol{k}|}+\frac{\boldsymbol{p_i}\cdot \boldsymbol{k}}{E_i^2|\boldsymbol{k}|^2}+\frac{(\boldsymbol{p_i}\cdot \boldsymbol{k})^2}{E_i^3|\boldsymbol{k}|^3}+\cdots\Big ),  i=1,2.
\eea
\bea
\frac{1}{\boldsymbol{q}^2}=\frac{1}{(E_1+E_2)^2}\Big (1-\frac{(\boldsymbol{p_1}+\boldsymbol{p_2})^2}{(E_1+E_2)^2}+\frac{2\boldsymbol{k}\cdot(\boldsymbol{p_1}+\boldsymbol{p_2})}{(E_1+E_2)^2}+\cdots \Big).
\eea

In order to get  analytical form of the approximate results, we factorize the integrand of $|\boldsymbol{p_1}|$ and $|\boldsymbol{p_2}|$ into the product of the steep and smooth parts,
\bea
&&\iint \frac{d^3\boldsymbol{p_1}}{(2\pi)^3}\frac{d^3\boldsymbol{p_2}}{(2\pi)^3} \frac{\Psi_1(\boldsymbol{p_1})\Psi_2(\boldsymbol{p_2})}{\sqrt{2E_1}\sqrt{2E_2}}{\cal M}_1^{\mu\nu}(\boldsymbol{p_1},\boldsymbol{p_2},\boldsymbol{k})\tilde{h}_{\mu\nu}(\boldsymbol{q})=\iiiint d|\boldsymbol{p_1}|d|\boldsymbol{p_2}|d\Omega_1 d\Omega_2  \nn \\
&&\sin \theta_1( \cos \phi_1 +i \sin \phi_1) \sin \theta_2( \cos \phi_2 +i \sin \phi_2)  I_1(|\boldsymbol{p_1}|,|\boldsymbol{p_2}|)I_2(\boldsymbol{p_1},\boldsymbol{p_2}),
\eea
where $I_1$ and $I_2$ are the steep and smooth functions of $|\boldsymbol{p_1}|$ and $|\boldsymbol{p_2}|$ respectively,
\bea
I_1&=&\frac{a_0^8|\boldsymbol{p_1}|^3|\boldsymbol{p_2}|^3}{(a_0^2|\boldsymbol{p_1}|^2+1/4)^3(a_0^2|\boldsymbol{p_2}|^2+1/4)^3}, \nn \\
I_2&=&-\frac{1}{a_0^3}\frac{1}{(2\pi)^5\sqrt{E_1E_2}}{\cal M}_1^{\mu\nu}(\boldsymbol{p_1},\boldsymbol{p_2},\boldsymbol{k})\tilde{h}_{\mu\nu}(\boldsymbol{q}).
\eea
The steep function $I_1$ is centered at $|\boldsymbol{p_1}|=|\boldsymbol{p_2}|=1/(\sqrt{2}a_0)$, thus we can make the approximation:
\be
I_1\approx (\frac{8}{3\pi})^2 \delta(|\boldsymbol{p_1}|-1/(\sqrt{2}a_0))\delta(|\boldsymbol{p_2}|-1/(\sqrt{2}a_0)),
\ee 
where $(8/3\pi)^2$ is the normalized constant. After calculating the integral, we finally get an analytical approximate form of the radiation power,
\begin{figure}
	\begin{center}
		\includegraphics[width=0.65\textwidth]{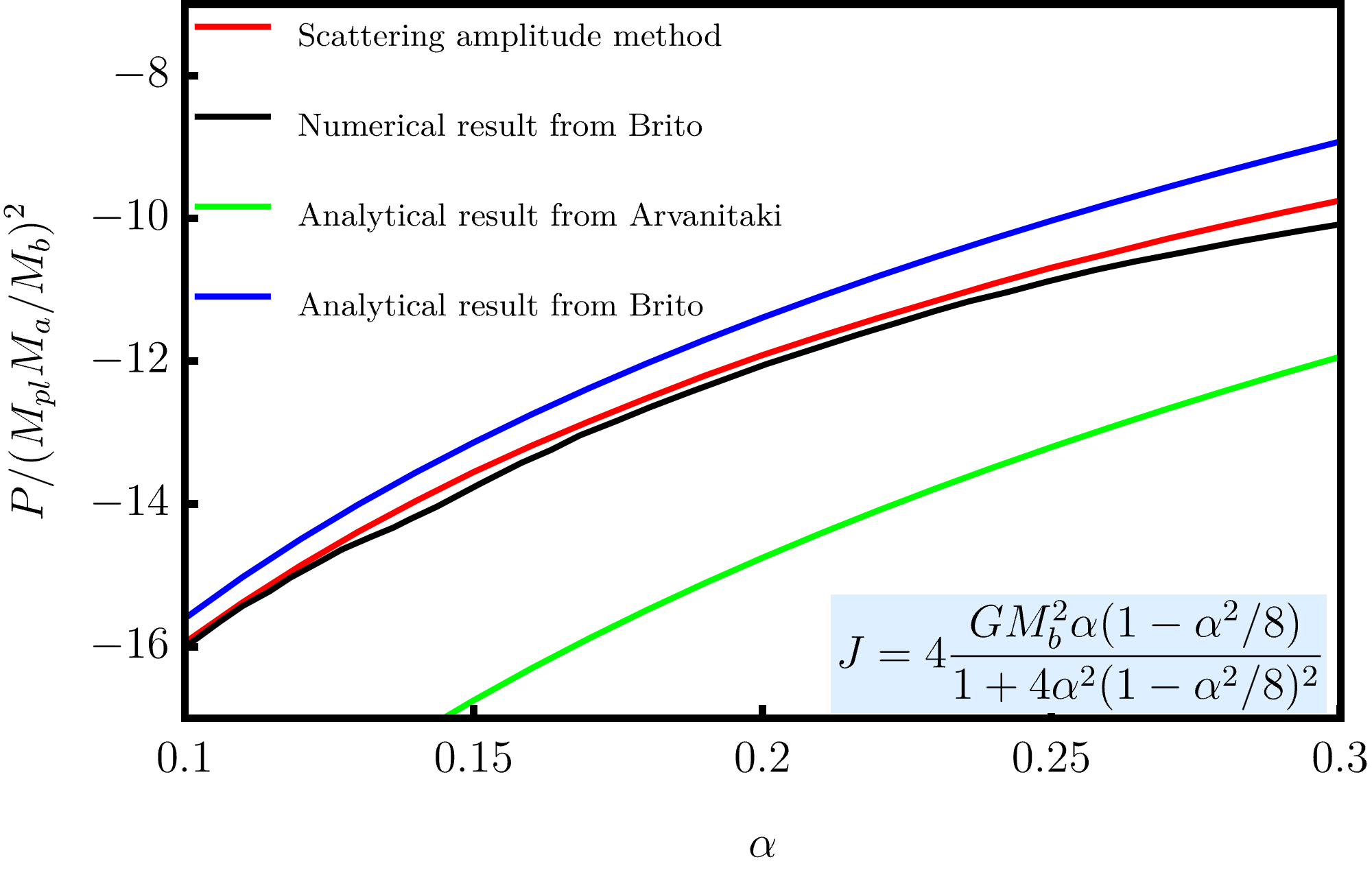}
		\caption{The GW radiation power with different “fine structure constant” in the Kerr BH's background field. The red line represents the numerical result through the scattering amplitude method in our study. We compare this result with the numerical result from Brito~\cite{Brito:2017zvb} (black line), the analytical results from Arvanitaki~\cite{Arvanitaki:2010sy} (green line), and Brito~\cite{Brito:2014wla} (blue line).}\label{p}
	\end{center}
\end{figure}
\bea \label{ana1}
P &=& \frac{2.436 \times 10^{-38} (M_a/\text{GeV})^{2}\alpha^{15}(2.977 \times 10^{38}-3.721 \times 10^{37}\alpha^{2})^{2}}{(M_b/\text{GeV})^6(2+\alpha^2)^{11}(4+\alpha^2)^4}\bigg{\{}(M_b/\text{GeV}) ^ { 4 }\\ \nn 
&&\bigg{[}\frac{448.0}{\alpha}+2.583 \times 10^{3} \alpha +6.828 \times 10^{3} \alpha^{3}+1.094 \times 10^{4} \alpha^{5}+1.182 \times 10^{4} \alpha^{7} \\ \nn 
&+&9.086 \times 10^{3} \alpha^{9}+5.092 \times 10^{3} \alpha^{11} 
+2.097 \times 10^{3} \alpha^{13}+629.4 \alpha^{15}+134.4  \alpha^{17}\\ \nn &+&19.36 \alpha^{19}+1.691 \alpha^{21}+0.068 \alpha^{23} \bigg{]}+J(M_b/\text{GeV})^2\sqrt{2+\alpha^2}\bigg{[}-1.257\times 10^{41}\\ \nn &-&6.915\times 10^{41} \alpha^{2}-1.746\times 10^{42} \alpha^{4} -2.675\times 10^{42} \alpha^{6}-2.771\times 10^{42} \alpha^{8} \\ \nn &-&2.047\times 10^{42} \alpha^{10} 
-1.105\times 10^{42} \alpha^{12}-4.392\times 10^{41} \alpha^{14}-1.277\times 10^{41} \alpha^{16}\\ \nn   
&-&2.648\times 10^{40} \alpha^{18}-3.719\times 10^{39} \alpha^{20}-3.176\times 10^{38} \alpha^{22}-1.248\times 10^{37} \alpha^{24}\bigg{]} \\ \nn 
&+&J^2\alpha\bigg{[}1.764\times 10^{79}+1.012\times 10^{80} \alpha^{2}+2.687\times 10^{80} \alpha^{4}+4.378\times 10^{80} \alpha^{6}\\ \nn 
&+&4.886\times 10^{80} \alpha^{8}+3.944\times 10^{80} \alpha^{10}+2.370\times 10^{80} \alpha^{12}+1.074\times 10^{80} \alpha^{14}\\ \nn 
&+&3.668\times 10^{79} \alpha^{16}+9.330\times 10^{78} \alpha^{18}+1.717\times 10^{78} \alpha^{20}+2.166\times 10^{77} \alpha^{22} \\ \nn 
&+& 1.677\times 10^{76} \alpha^{24}
+6.023\times 10^{74} \alpha^{26}\bigg{]}\bigg{\}}\text{GeV}^2,
\eea
where $M_a$ is the mass of the axion cloud and we take the number of axions in the $(n=2,l=1,m=1)$ state as $N\approx M_a/m_a$.
Since $\alpha<1$, very higher-order terms of $\alpha$ might make negligible contributions. Considering the above calculation is valid and accurate mostly for $\alpha\lesssim 0.5$, we firstly find the largest terms in the three square brackets of Eq.~\ref{ana1} at $\alpha=0.5$. Then we keep the dominant terms and omit the terms which are one or more orders of magnitude smaller than the largest terms. Thus we get the further approximated form from Eq.~\eqref{ana1},
\bea
P&=&\frac{(M_a/\text{GeV})^{2}\alpha^{14}}{(M_b/\text{GeV})^6(2+\alpha^2)^{11}(4+\alpha^2)^4}\label{ana2} \bigg{[}(M_b/\text{GeV})^4(9.671\times 10^{41}+5.577\times 10^{42}\alpha^{2} \\ \nn 
&+& 1.474\times 10^{43}\alpha^4+2.361\times 10^{43} \alpha^{6}) +J(M_b/\text{GeV})^2 \alpha(-3.839\times 10^{80} \\ \nn
&-&2.111\times 10^{81}\alpha^2-5.329\times 10^{81}\alpha^{4}-8.165\times 10^{81}\alpha^{8}) 
+J^2\alpha^2(3.809\times 10^{118} \\ \nn 
&+&2.184\times 10^{119}\alpha^2+5.799\times 10^{119}\alpha^{4}+9.450\times 10^{119}\alpha^{6})\bigg{]}~\text{GeV}^2.
\eea

\begin{figure}[t]
	\begin{center} 
		\includegraphics[width=0.48\textwidth]{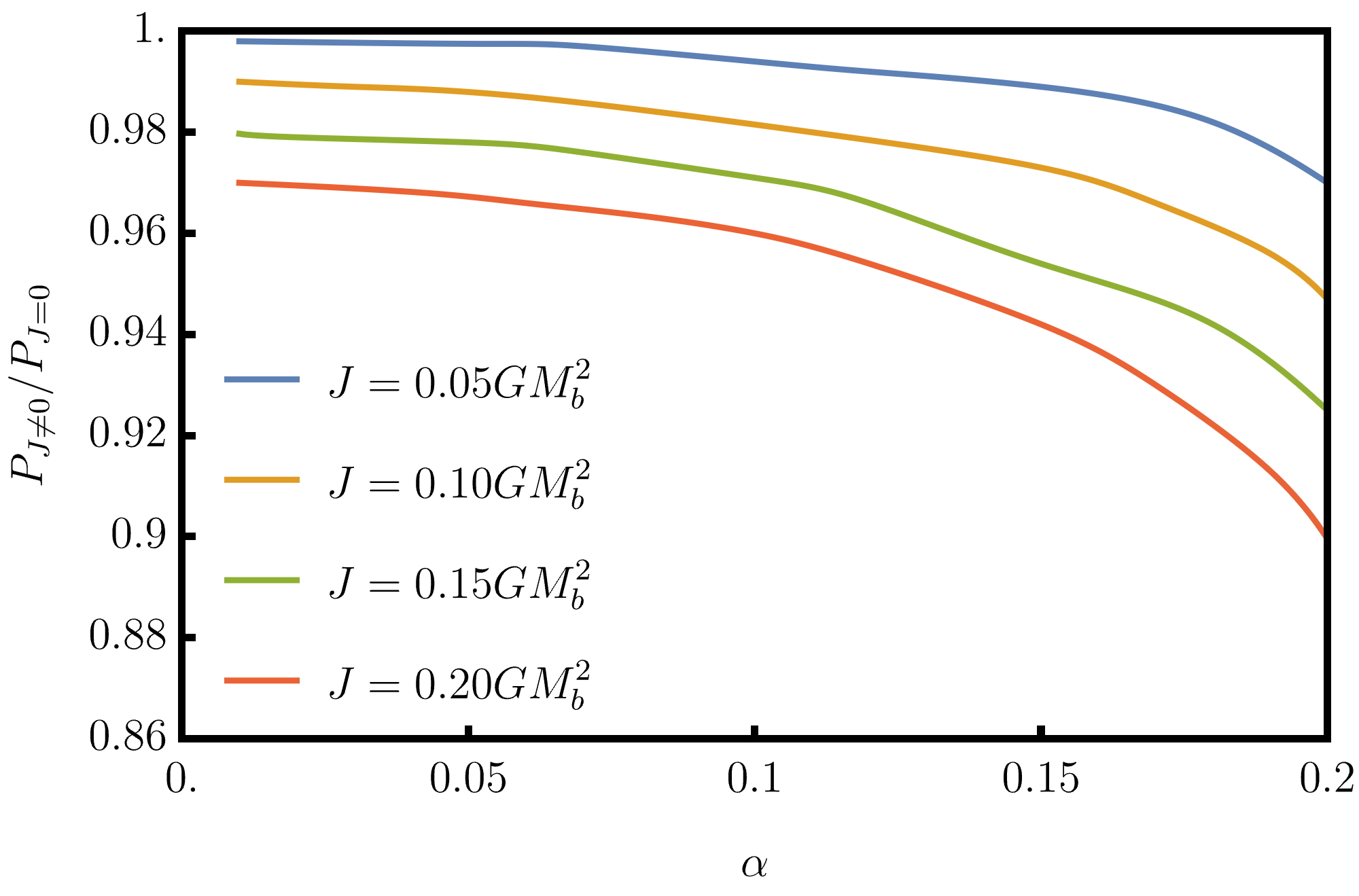}
		\includegraphics[width=0.48\textwidth]{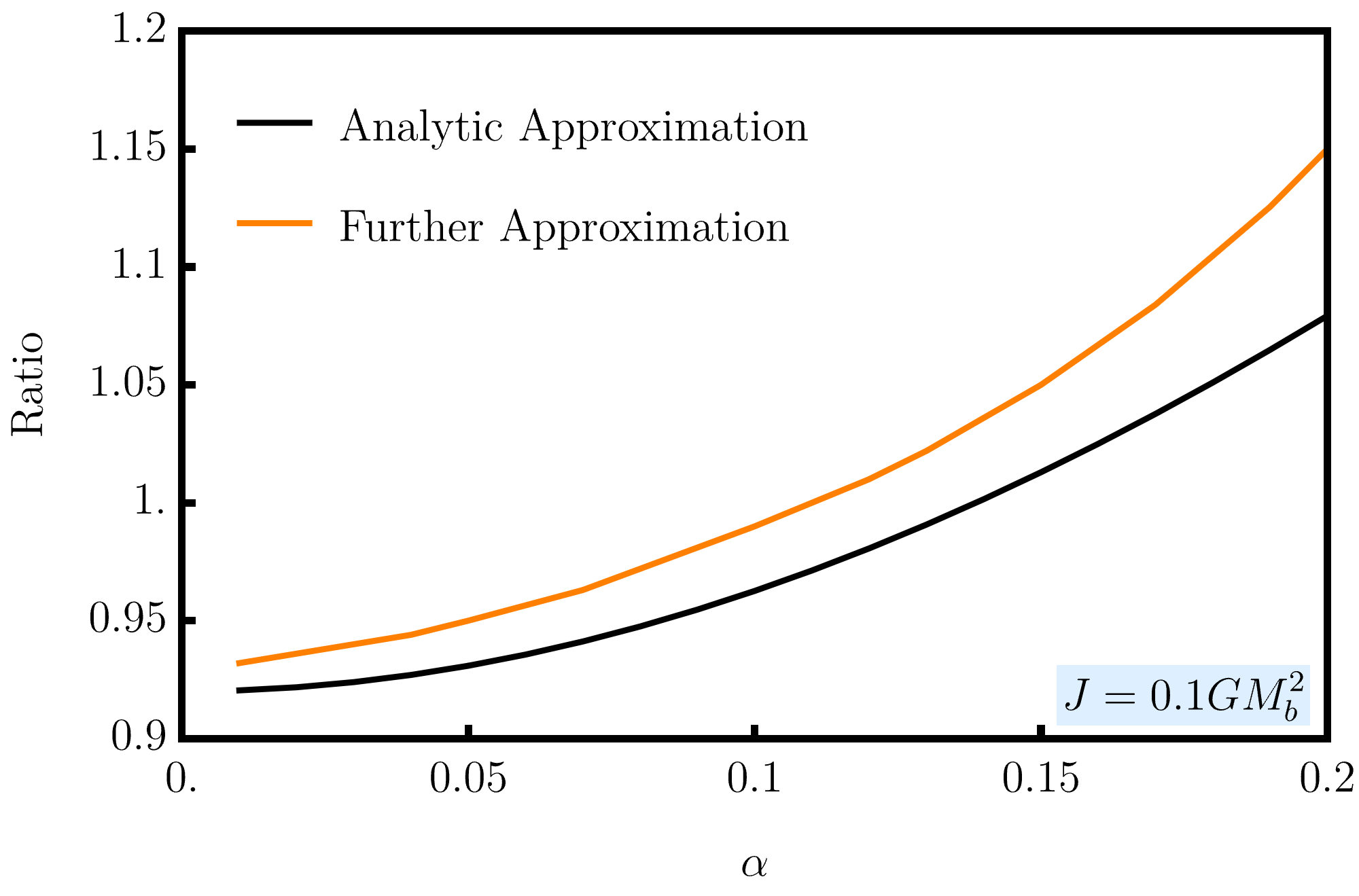}
		\caption{Left: The ratio of GW radiation power with different spin of BH $J$ to the spinless case $J=0$. The blue, orange, green, red lines correspond to $J=$0.05,~0.10,~0.15,~0.20$GM_b^2$ respectively. Right: The ratio of analytical approximate results in Eqs.~\eqref{ana1} (black) and \eqref{ana2} (orange) to numerical result at $J=0.1GM_b^2$.   } \label{j}
	\end{center}
\end{figure}

We also calculate the numerical result of the multiple integrals over $\boldsymbol{p_1}$,$\boldsymbol{p_2}$, and $\boldsymbol{k}$ in Eq.~\eqref{P} through the Monte Carlo method to obtain the radiation power. The GW radiation power with different “fine structure constant” is shown in Fig.~\ref{p} where we take BH spin $J=4\frac{GM_b^2\alpha(1-\alpha^2/8)}{1+4\alpha^2(1-\alpha^2/8)^2}$ as in Ref.~\cite{Brito:2017zvb}. We compare the result through the scattering amplitude method with the traditional method from Brito~\cite{Brito:2017zvb}\cite{Brito:2014wla} and Arvanitaki~\cite{Arvanitaki:2010sy}, our result is a good approximation to numerical result from Brito~\cite{Brito:2017zvb} at small $\alpha$ as expected.
The analytic result derived from Brito~\cite{Brito:2014wla} is obtained in the background of spinless BH which correspond to take $\tilde{h}_{\mu\nu,J}=0$ [Eq.~\eqref{hqj}] approximation in the scattering amplitude method. While the analytical result from Arvanitaki~\cite{Arvanitaki:2010sy} is around two to five orders of magnitude smaller than the numerical result since it is calculated in the flat space. In fact, the formula used in the traditional method in the flat space background is equivalent to the formula of  radiation power generated by the classical source. We postpone the detailed discussion in Sec.~\ref{dis}.

 The ratio of the magnitude of different BH spin angular momentum $J$ to the spinless case $J=0$ is shown in the left panel of Fig.~\ref{j}. We can find that the radiation power decreases with the increase of the BH spin angular momentum when the spin is small. To verify the validity of Eqs.~\eqref{ana1} and \eqref{ana2}, we also plot the ratio of the analytical approximate results to the numerical result where we take $J=0.1GM_b^2$ in the right panel of Fig.~\ref{j}. They are good approximations within the range of $\alpha$ and $J$  considered in this work. 

\section{Discussion} \label{dis}

We note that the traditional method used in Ref.~\cite{Arvanitaki:2010sy} is based on the formula for the GW radiation power,
\be
\frac{dP}{d\Omega}=\frac{G\omega^2}{\pi}T_{ij}^{TT*}(\omega,\boldsymbol{k})T_{ij}^{TT}(\omega,\boldsymbol{k}),
\ee
where 
\be
T_{ij}(\omega,k)\equiv \int d^3\boldsymbol{x}T_{ij}(\omega,\boldsymbol{x})e^{-i\boldsymbol{k}\cdotp\boldsymbol{x}},
\ee
with $|k|=\omega$ and the superscript $TT$ means that we take the transverse-traceless part,
\be
T_{ij}^{TT}\equiv (P_{ii'}P_{jj'}-\frac{1}{2}P_{ij}P_{i'j'})T_{i'j'},
\ee
where
\be
P_{ij}=\delta_{ij}-\frac{k_ik_j}{\boldsymbol{k}^2} .
\ee

\begin{figure}[H]
	\begin{center}
		\includegraphics[width=0.6\textwidth]{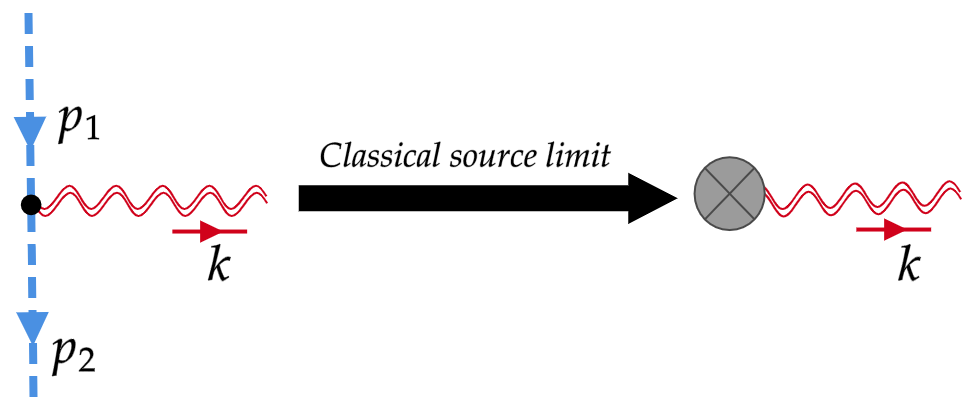}
		\caption{The Feynman diagram for scattering amplitude of $|\boldsymbol{p_1},\boldsymbol{p_2}\rangle \rightarrow |\boldsymbol{k}\rangle $ process in classical source limit.  } \label{tra}
	\end{center}
\end{figure}

We now derive this formula from field theory in classical limit. Since this formula is applicable in both transition and annihilation processes, we just derive it from transition process for simplicity. We can consider a classical source emits a graviton as a transition process $|p_1\rangle \rightarrow |p_2,k\rangle $ where $k$ is the momentum of emitted graviton, and $p_1$, $p_2$ are the momenta of the source in initial and final states. The decay width can be calculated similarly as  Eq.~\eqref{gamma},
\be
\Gamma \simeq \frac{1}{2M_j}\int\frac{d^3\boldsymbol{k}}{(2\pi)^3}\frac{1}{2|\boldsymbol{k}|}\frac{1}{2M_j}\big{|}{\cal M}(p_1\rightarrow p_2,k)|^2 2\pi \delta(E_1-E_2-|\boldsymbol{k}|),
\ee
where $M_j$ is the mass of the classical source. And the scattering amplitude ${\cal M}(p_1\rightarrow p_2,k)$ is just that we replace the propagator of graviton and $M_b$ with polarization tensor of physical graviton and $M_j$ respectively in $-i{\cal M}_2^{\mu\nu}$. As shown in Fig.~\ref{tra}, in classical source limit,
\be
{\cal M}(p_1\rightarrow p_2,k)=-i\kappa M_j T_{\mu\nu}(\omega,\boldsymbol{k})\epsilon^{\mu\nu},
\ee
where $\omega=|\boldsymbol{k}|$, then we get
\be
\Gamma=\sum_{s=\pm}\int\frac{d\omega d\Omega \omega \kappa^2}{32\pi^2}|T_{\mu\nu}(\omega,\boldsymbol{k})\epsilon^{s,\mu\nu}|^2 \delta(E_1-E_2-\omega),
\ee
the polarization sum $\sum_{s_1,s_2=\pm}\epsilon^{s_1,\mu\nu*}\epsilon^{s_2,\alpha\beta}$can be calculated by using Eq.~\eqref{epsilon}. Finally we get
\be
\Gamma=\int  d\Omega \frac{G\omega_{12}}{\pi}T_{ij}^{TT*}(\omega_{12},\boldsymbol{k})T_{ij}^{TT}(\omega_{12},\boldsymbol{k}),
\ee
where $\omega_{12}=E_1-E_2$, thus the radiation power is
\be
\frac{dP}{d\Omega}=  \frac{G\omega_{12}^2}{\pi}T_{ij}^{TT*}(\omega_{12},\boldsymbol{k})T_{ij}^{TT}(\omega_{12},\boldsymbol{k}).
\ee
This formula is only applicable in classical source limit in flat space. But for axions, process $|p_1\rangle \rightarrow |p_2,k\rangle $ is kinematically forbidden due to the on-shell requirement in flat space. Thus the flat space approximation in Ref.~\cite{Arvanitaki:2010sy} is equivalent to the classical source limit in which the matter has no on-shell requirement. For the real physical process, we must consider the background field effect. We need to take Kerr BH into account for this process as the Feynman diagrams in Fig.~\ref{fd} and then treat it as a classical source. This is the method we have used in this work.

\section{conclusion}\label{conc}

Using the method of scattering amplitude, we have calculated the GW radiation power of the axions  annihilation process in the axion cloud around Kerr BH. Compared with the traditional method for calculating the GW radiation power, the scattering amplitude method has the following advantages:

1. The calculation is simple and straightforward. The Hamiltonian of the traditional method is $H=H_0+H_I$ with $H_0=H_{0,a}+H_{bf,a},H_I=H_{int,a}$, where $H_{0,a}$ and $H_{int,a}$ represent the free Hamiltonian and gravitational self-interaction of axion field. $H_{bf,a}$ represents the interaction between the background and axion field. The traditional method is equivalent to deriving the eigenstates of $H_0$ first and then calculating the transition rate of different eigenstates under perturbation $H_I$. While for the method of scattering amplitude, we take the form of Hamiltonian as $H_0=H_{0,a}+H_{0,b}$, $H_I=H_{a,b}+H_{int,a}$, where $H_{0,b}$ is the free Hamiltonian of BH and $H_{a,b}$ represents the interaction between Kerr BH and axion particle. Although the perturbation method of scattering amplitude is valid for small $\alpha$, the numerical calculation is much easier than the traditional method. 
  
2. It is natural and easy to include the Kerr metric effects. We can divide the calculation of scattering amplitude into two parts as in Eq.~\eqref{sc}, where $M_{1\mu\nu}$ stands for axions annihilation process and $M_{2\mu\nu}$ corresponds to metric effects induced by Kerr BH. The metric effects of Kerr BH act as a background field $\tilde{h}_{\mu\nu}=\tilde{h}_{\mu\nu,0}+\tilde{h}_{\mu\nu,J}$ which is included in $M_{2\mu\nu}$ as in Eq.~\eqref{M21}. Since the scattering method is valid for small $\alpha$, we only expand the metric up to the first nonvanish terms in the weak-field approximation. And note that only the first-order term of BH spin $J$ in Eq.~\eqref{hj} is retained. In principle,  when the spin of Kerr BH is large, higher power terms of spin $J$ need to be included in Eq.~\eqref{hj} for a more accurate result and we can get these terms by expanding the Kerr metric directly. For simplicity, we only retain linear terms of $J$ which is a good approximation for the small BH spin.

3. The microscopic physical process is intuitive. The annihilation process of two axions can be considered as a decay process of a three-body bound state.  For this decay process, there are four Feynman diagrams corresponding to different scattering channels. We can naturally extend our calculation to four, six, or more axions annihilation processes.

4. It is clear and simple to demonstrate the analytic approximation formulas. Since the wave function of the axion state $(n=2,l=1,m=1)$ is centered at $p\sim \alpha m_a/2$, we can get an analytic approximate formula with the spin effect from the phase integration of scattering amplitude.  While the spin effect is hard to be included in the analytic approximation in the traditional method.

We have cross-checked and confirmed the numerical results from the traditional method~\cite{Brito:2017zvb} which is important for the GW experiments and axion search. More precise calculations, implications for GW or axion detections, and more broad applications are working in progress.
\begin{acknowledgments}
This work was supported by the National Natural Science Foundation of China (NNSFC) under Grant No. 12205387 and Guangdong Major Project of Basic and Applied Basic Research (Grant No. 2019B030302001).
\end{acknowledgments}


\end{document}